\documentclass[preprint,showpacs,preprintnumbers,amsmath,amssymb, superscriptaddress,longbibliography,nofootinbib]{revtex4-1}

\usepackage{textcomp}
\usepackage{makeidx}
\usepackage{amsmath}
\usepackage{subfigure}
\usepackage{amssymb}
\usepackage{hyperref}
\usepackage{graphicx}
\usepackage{epstopdf}
\usepackage{color}
\usepackage{soul}

\begin{document}

\title{A new exact solution of black-strings-like with a dS core}

\author{Milko Estrada}
\email{milko.estrada@gmail.com}
\affiliation{Departmento de f\'isica, Universidad de Antofagasta, 1240000 Antofagasta, Chile}

\date{\today}

\begin{abstract}
We provide a new five dimensional black string--like solution by means of the embedding of a four dimensional regular black hole into a compact extra dimension. We enunciate a list of constraints in order to that the five dimensional black string--like solution to be regular, and, following these constraints we construct our solution. Instead of the usual singularity, it is formed a core whose topology corresponds to the product between the de--Sitter core of the four dimensional Hayward solution and $S^1$. The horizon has topology $S^2 \times S^1$. At infinity of the radial coordinate the regular four dimensional geometry is asymptotically flat, {\it i.e}, at this place the topology of the complete solution corresponds to the product between Minkowski and $S^1$. At the induced four dimensional geometry we compute the correct values of temperature and entropy.
\end{abstract}

\maketitle

\section{Introduction}
In the last decades, several branches of theoretical physics have predicted the existence of extra dimensions, examples of this latter are the string theory, brane world models, etc. In this context, the study of higher dimensional black hole solutions have showed some different features respect to the well known four dimensional solutions. Examples of this latter are higher dimensional black holes in modified theories of gravity, black holes in brane world models, black strings (branes) and black rings \cite{Emparan:2008eg}.     

Black string are the simplest extension of the black hole solutions. The simplest version of black string is given by the line element :
\begin{equation} \label{Kalusa}
    ds_{5D}^2=d{s}_{4D}^2+ dz^2,
\end{equation}
 where $d{s}_{4D}^2$ corresponds to the four dimensional Schwarzschild space time. The line element \eqref{Kalusa} corresponds to the Kaluza-Klein black string \cite{Gregory:2008rf}. So, the last equation corresponds to a solution of the Einstein field equations $G_{M N}=0$ and it was obtained by the oxidation of the Schwarzschild solution \cite{Cisterna:2018mww}. This solution has $S^2 \times R$ horizon topology for $z$ non compact, {\it i.e} has a {\it cylindrical event horizon} \cite{Gregory:2008rf}. If the extra coordinate $z$ is compact the topology of the horizon is $S^2 \times S^1$. The horizon topology differs from the usual $S^3$ topology of the five dimensional Schwarzschild-Tangherlini solution \cite{Kol:2002xz}.
 On the other hand, the topology of the central singularity corresponds to the product between the Schwarzschild singularity and $R$ ($S^1$) for a non compact (compact) extra coordinate.
 
Black string have been widely studied in literature for different contexts. For example solutions in GR with cosmological constant \cite{Cisterna:2017qrb}, with axionic scalar fields in  Horndeski gravity \cite{Cisterna:2017jmv}, in $f(R)$ gravity \cite{Sheykhi:2013yga}, Lovelock gravity \cite{Giacomini:2018sho} and in brane world models \cite{Chamblin:1999by,Kanti:2002fx,Nakas:2019rod}. Other examples of recent applications in references \cite{Nakas:2020crd,Hosseini:2020vgl,Sheykhi:2020fqf}.

Black strings with a central singularity are easily constructed by embedding singular black hole solutions into an extra dimension. However, in the literature there are also non singular black holes solutions that could be embedded into an extra dimension, which clearly would not lead to a central singularity. These latter solutions are known as {\it regular black holes}. In these models, instead the formation of a central singularity, it is formed a {\it de--Sitter core}, {\it i.e}, the solution behaves as a de--Sitter space time near the origin. Thus, all the curvature invariants are regular everywhere. One very interesting and intriguing model is the Hayward metric \cite{Hayward:2005gi}. Using this metric, the formation of the Sitter core is associated to quantum fluctuations, where the energy density is of order of Planck units near the origin. This latter model is called Planck stars \cite{Rovelli:2014cta,DeLorenzo:2014pta}. Although these models are theoretical, in reference \cite{Rovelli:2017zoa}, using radio astronomy data, is conjetured that Planck Stars represent a speculative but realistic possibility of testing quantum gravity effects. 

Thus, motivated by the number of applications of black strings, where are embedded singular black holes into an extra dimension, is of physical interest the embedding of regular black hole solutions also into an extra dimension. In this work, we will study this latter mentioned problem. For this, we will use the Hayward metric as toy model. We will compute the five dimensional solution and analyze the way how this solution modifies locally the four dimensional geometry and the information of the energy momentum tensor.  
Furthermore, we will discuss the first law of thermodynamics at the induced four dimensional geometry under some assumptions below discussed.   

\section{Our Model}
The Einstein field equation in $5D$ are given by:
\begin{equation} \label{EcuacionDeEinstein}
    G^M_{\,\,\,N}+\Lambda_{5D} \delta^M_{\,\,\,N}= 8 \pi \tilde{G}_{5D} T^M_{\,\,\,N},
\end{equation}
where $M,N=0,1,2,3,5$, and where $\tilde{G}_{5D}$ is the five dimensional Newton constant. For simplicity, we consider arbitrarily that this constant has a magnitude equal to unity.

We will study the following space time:
\begin{align}
     ds_{5D}^2=& W(z) \cdot d\bar{s}_{4D}^2+ dz^2 \label{elementodelinea} \\
              =& \exp(-2A(z)) \cdot g_{\mu \nu}^{(4D)} dx^\mu dx^\nu + dz^2 \label{elementodelineaIntermedio} \\
              =& \exp(-2A(z)) \cdot \left (-f(r)dt^2+f(r)^{-1}dr^2+r^2 d\Omega^2 \right)+ dz^2 \label{Elementodelinea1}
\end{align}
with $\mu,\nu=0,1,2,3$ and where  the warp factor (W) is a function that depend only on the compact extra coordinate $z$. Thus, the $4D$ geometry, $d\bar{s}_{4D}^2$, is deformed by the warp factor, and thus, our solution corresponds to a {\it non-uniform string-like} \cite{Kol:2002xz}. The energy momentum tensor is:
\begin{equation}
    T^M_{\,\,\,N}=\mbox{diag}(-\rho,p_r,p_\theta,p_\phi,p_z),
    \end{equation}
where $\rho,p_r,p_\theta,p_\phi,p_z$ represent the energy density and the $r,\theta,\phi,z$ pressure components. We will impose that the energy momentum tensor has the form:
    \begin{align}
       T^M_{\,\,\,N} =& f \big ( W  (z) \big) \cdot \mbox{diag}(-\bar{\rho}(r),\bar{p}_r(r),\bar{p}_\theta(r),\bar{p}_\phi(r),\tilde{p}_z(r)),
\end{align}
where $f \big ( W  (z) \big)$ represents a function of the warp factor. For convenience, we will write the energy momentum tensor as follow:
\begin{equation} \label{EnergiaMomentum}
    T^M_{\,\,\,N}=\delta^M_\mu \delta^\nu_N \cdot f \big ( W  (z) \big) \cdot \bar{T}^\mu_{\,\,\,\nu} + T^5_{\,\,\,5},
\end{equation}
 where $\bar{T}^\mu_{\,\,\,\nu}$ represents to the energy momentum tensor associated with the four dimensional geometry:
\begin{equation} \label{EnergiaMomentum4D}
    \bar{T}^\mu_{\,\,\,\nu}=   \mbox{diag}(-\bar{\rho}(r),\bar{p}_r(r),\bar{p}_\theta(r),\bar{p}_\phi(r))
\end{equation}

Thus $\bar{\rho}(r),\bar{p}_r(r),\bar{p}_\theta(r),\bar{p}_\phi(r)$ represent the energy density and the $r,\theta,\phi$ pressure components of the four dimensional geometry. On the other hand $T^5_{\,\,\,5}$ corresponds to the energy momentum tensor along the extra dimension:
\begin{equation} \label{EnergiaMomentum5}
    T^5_{\,\,\,5}= \delta^5_5  f \big ( W  (z) \big) \tilde{p}_z(r).
\end{equation}
where $\tilde{p}_z(r)$ represents the radial dependent part of $T^5_{\,\,\,5}$.

Thus, we can point out that the warp factor locally modifies both the geometry as the content of the energy momentum tensor. Examples in $5D$, where the energy momentum tensor depends on the extra coordinate we can see in reference \cite{Kennedy:2000vq} for brane world models, and, where the energy momentum tensor depends on the radial coordinate in reference \cite{Ali:2019mxs} for black strings surrounded by quintessence matter.  

\subsection{Constraints for regularity of the solution} \label{restricciones}

For the line element \eqref{Elementodelinea1}, the corresponding expressions for the Ricci scalar is:
\begin{align} \label{Ricci5D}
    R_{5D}=-20 (\dot{A})^2+8 \ddot{A}+ W^{-1} \bar{R}_{4D},
\end{align}
where { \bf the dot indicates derivation with respect to the extra coordinate ($z$)}, and where  $\bar{R}_{4D}$ corresponds to the Ricci scalar of the four dimensional geometry whose line element is $d\bar{s}_{4D}^2$ in equation \eqref{elementodelinea}:
\begin{equation} \label{Ricci4D}
    \bar{R}_{4D}= -f''- \frac{4}{r}f'+ \frac{2}{r^2} \left ( 1-f \right) .
\end{equation}
where $'$ represents derivation with respect to the radial coordinate ($r)$. 

On the other hand, the Kretschmann scalar is given by:
\begin{align} \label{Kre5D}
    K_{5D}= 40 (\dot{A})^4+16 (\ddot{A})^2-32 \ddot{A} (\dot{A})^2- 4 W^{-1} (\dot{A})^2 \bar{R}_{4D} + W^{-2} \bar{K}_{4D},
\end{align}
where the four dimensional Kretschmann scalar $\bar{K}_{4D}$, which also corresponds to the $4D$ geometry $d\bar{s}_{4D}^2$, is given by:
\begin{equation} \label{4D}
   \bar{K}_{4D}=(f'')^2+ \frac{4}{r^2} (f')^2 + \frac{4}{r^4} (1-f)^2 .
\end{equation}

Thus, in order to describe a regular solution, {\it i.e} both invariants $R_{5D}$ and $K_{5D}$ must be regular, we will impose the following constraints: 
\begin{itemize}
    \item The function $A(z)$ must be continuous in its first and second derivatives. So, our model does not have $\mathcal{Z}_2$ symmetry as brane world models \cite{Randall:1999ee,Randall:1999vf,Aros:2012xi}, because the second derivative of the absolute value behaves as Dirac delta and is not regular in $z=0+N\pi$, with $z$ compact.
    \item The function $A(z)$ must be such that the Warp factor $W \left ( A(z) \right)$ can not be zero.
    \item The four dimensional geometry must be regular, therefore the four dimensional invariants, $\bar{R}_{4D}$ and $\bar{K}_{4D}$ must be regular. 
\end{itemize}

\subsection{Equations of motion}

We will use the following ansatz for the $f(r)$ function:
\begin{equation} \label{funcionf}
    f(r)=1-\frac{2m(r)}{r},
\end{equation}
where $m(r)$ is the so called mass function and, as it was above mentioned, must be such that the four dimensional invariants must be regular.

For the line element \eqref{Elementodelinea1} with the ansatz \eqref{funcionf}, using the coinstraints $T^0_0=T^1_1$ and $T^2_2=T^3_3$, the components of the Einstein equations \eqref{EcuacionDeEinstein} are:

$(t-t)$ and $(r-r)$ components:
\begin{equation} \label{tt}
    6(\dot{A})^2-3 \ddot{A} + \Lambda_{5D}-\frac{2}{r^2}W^{-1} \frac{dm}{dr}=- 8 \pi f(W) \bar{\rho}
\end{equation}
$(\theta-\theta)$ and $(\phi-\phi)$ components:
\begin{equation} \label{thetatheta}
    6(\dot{A})^2-3 \ddot{A} + \Lambda_{5D}-\frac{1}{r}W^{-1} \frac{d^2m}{dr^2}= 8 \pi f(W) \bar{p}_\theta
\end{equation}
and the $(z-z)$ component:
\begin{equation} \label{zz}
    6(\dot{A})^2 + \Lambda_{5D}-\frac{2}{r^2}W^{-1} \frac{dm}{dr}-\frac{1}{r}W^{-1} \frac{d^2m}{dr^2}= 8 \pi f(W) \tilde{p}_5 .
\end{equation}

For solving these equations we will use the following steps:
\begin{itemize}
    \item In the $(z-z)$ component we assume that
    \begin{equation} \label{EcuacionA}
        6(\dot{A})^2 =- \Lambda_{5D}.
    \end{equation}

From this latter equation it is direct to check that $\ddot{A}=0$. Thus, using the condition \eqref{EcuacionA}, the two first terms of the equation \eqref{zz} and the three first terms of the equations \eqref{tt} and \eqref{thetatheta} are equal to zero. 

Thus, for determining the form of $A(z)$, solving the equation \eqref{EcuacionA} is enough. Below we will discuss about the solution.

\item From equation \eqref{tt} it is direct to check that:
\begin{equation}
    \frac{2}{r^2}W^{-1} \frac{dm}{dr}= 8 \pi f(W) \bar{\rho},
\end{equation}
thus, it is fulfilled that:
\begin{equation} \label{EcuacionW}
    f \left(W (z) \right) = W^{-1}= \exp(2A(z))
\end{equation}
and
\begin{equation} \label{funcionDeMasa}
    m(r)= 4 \pi \int_0^r x^2 \bar{\rho} (x) dx,
\end{equation}

The equation \eqref{funcionDeMasa} corresponds to the usual mass function for regular black holes \cite{Aros:2019auf,Aros:2019quj}. Below will be described the form for the four dimensional energy density $\bar{\rho}$.
\item From equation \eqref{thetatheta} it is direct to check that:
\begin{equation} \label{presion4DTheta}
-\frac{1}{ r} \frac{d^2m}{dr^2}= 8 \pi \bar{p}_\theta
\end{equation}
\item The radial and tangential pressures are easily computed by the conditions $T^0_0=T^1_1$, {\it i.e} $-\bar{\rho}=\bar{p}_r$ , and $T^2_2=T^3_3$, {\it i.e} $\bar{p}_\theta=\bar{p}_\phi$.
\item From equation \eqref{zz} it is direct to check that:
\begin{equation} \label{p5}
    -\bar{\rho}+\bar{p}_\theta=\tilde{p}_5.
\end{equation}
\end{itemize}

From equation \eqref{EcuacionW} we can see that the warp factor modifies the four dimensional geometry \eqref{elementodelinea}. Furthermore, from equations \eqref{EcuacionW} and \eqref{EnergiaMomentum}, we note that the warp factor also modifies the information of the four dimensional energy momentum tensor \eqref{EnergiaMomentum4D} and the energy momentum tensor along the extra dimension \eqref{EnergiaMomentum5}.

\subsection{The four dimensional geometry}
We will choose the Hayward \cite{Hayward:2005gi} metric as example. This model has a de--Sitter core, and thus, both the Ricci and the Kretschmann are regular everywhere. The energy density is given by \cite{DeLorenzo:2014pta}:
\begin{equation}
    \bar{\rho}= \frac{3}{2\pi} \frac{LM^2}{(2LM+r^3)^2},
\end{equation}
where $L$ is a constant parameter. Replacing into equation \eqref{funcionDeMasa}:
\begin{equation} \label{funciondeMasa1}
    m(r)=\frac{Mr^3}{2LM+r^3},
\end{equation}
where the $M$ parameter corresponds to the total mass \cite{Aros:2019quj}. The mass function behaves as $m(r) |_{r \approx 0} \approx (1/2L) r^3$ near the origin. Thus, the four dimensional solution \eqref{funcionf} behaves as a de--Siter space time near the origin, {\it i.e.} the solution has a de--Sitter core.
A complete analysis of the solution can be viewed in reference \cite{DeLorenzo:2014pta}. See also \cite{Aros:2019auf,Estrada:2020ptc} for a thermodynamics analysis .

It is worth to stress that both the energy density and the mass function ensure a well asymptotic behavior of the four dimensional solution \cite{Estrada:2020tbz}:
\begin{align}
        \displaystyle \lim_{r \to \infty} \bar{\rho} &=0 \label{AsintotaRho} \\
        \displaystyle \lim_{r \to \infty} m(r) &= \mbox{constant}=M. \label{AsintotaM}
    \end{align}
    
From equation \eqref{AsintotaM}, it is easy to check that in equation \eqref{presion4DTheta}:
\begin{equation} \label{AsintotaPr}
\displaystyle \lim_{r \to \infty} \bar{p}_\theta=0
\end{equation}

Thus, the four dimensional solution is asymptotically flat. Below we will discuss the consequences of this latter in the complete five dimensional solution. 

\subsection{Complete five dimensional solution}

We choose the following cosmological constant:
\begin{equation}
    \Lambda_{5D}=-\frac{6}{l^2},
\end{equation}
where $l$ corresponds to the AdS radius when the space time represents to an AdS space. From equation \eqref{EcuacionA} it is easy to check that the line element \eqref{elementodelinea} is:
\begin{equation}
     ds_{5D}^2= C \exp \left (\pm \frac{2}{l}z \right) \cdot d\bar{s}_{4D}^2+ dz^2.
\end{equation}
where $d\bar{s}_{4D}^2$ is given by the Hayward space time described in the previous subsection and $C$ is a positive constant.

As it was above mentioned for the regularity of the solution, the warp factor $W= C \exp \left (\pm \frac{2}{l}z \right)$ must be non zero. So, we need that the extra coordinate, $z$, to be compact such that $z = z + 2 N \pi$. This latter is because, for the positive (negative) branch, the warp factor tends to zero at $-\infty$ ( $\infty$).

The warp factor differs from the brane world models \cite{Randall:1999ee,Randall:1999vf} because our solution does not have $\mathcal{Z}_2$ symmetry. This latter is because in models with $\mathcal{Z}_2$ symmetry the second derivative $\ddot{A}$ behave as Dirac Delta, {\it i.e} are singular for $z=0+N\pi$, with $z$ compact, see equations \eqref{Ricci5D} and \eqref{Kre5D}. Although the four dimensional solution behaves asymptotically as the Schwarzschild space time, at infinity (of the radial coordinate) our solution does not coincide with the solution of reference \cite{Kanti:2002fx} , since our solution does not have $\mathcal{Z}_2$ symmetry and in our case the extra coordinate $z$ is compact.

As it was above mentioned, the four dimensional solution is regular. Thus, instead the formation of a black string where the radial singularity is represented by the product between the Schwarzschild singularity and $S^1$, it is formed a black string--like where the core is represented by the product between the de--Sitter core of the four dimensional Hayward solution and $S^1$. 

From equations \eqref{AsintotaRho} and \eqref{AsintotaPr} it is direct to check that in equation \eqref{p5}:
\begin{equation}
    \displaystyle \lim_{r \to \infty} \tilde{p}_5=0.
\end{equation}

So, the energy momentum tensor tends to zero at the asymptote of the radial coordinate. Thus, due that the four dimensional solution is asymptotically flat, our complete solution behaves at this place as an Anti de--Sitter five dimensional space time. On the other hand, also at the infinity of the radial coordinate the topology corresponds to the product between Minkowski and $S^1$. Furthermore at this asymptote, for zero cosmological constant, $l \to \infty$, our solution behaves as the Kaluza-Klein black string \cite{Gregory:2008rf}.

\section{Thermodynamics Analysis}

It is direct to check that for an embedding $z=z_0$, where $z_0$ represents an arbitrary location $z_0 \in z$, the induced metric is given by \cite{Aros:2012xi}:
\begin{equation} \label{metricIanducida}
    h_{\mu \nu} = g_{\mu \nu}^{(4D)} W(z_0) .
\end{equation}

In order to compute the first law of thermodynamics we will use the conditions $h_{tt}(r_+,M,z_0)=0$ and $\delta h_{tt}(r_+,M,z_0)=0$, where $h_{tt}$ and $r_+$ represent the temporal component of the induced metric and the horizon radius. These latter conditions can be viewed as constraints on the evolution along the space parameters \cite{Aros:2019auf,Zeng:2019huf}. From equations \eqref{metricIanducida} and \eqref{Elementodelinea1}, the temporal component of the induced metric is:
\begin{equation}
    h_{tt}=-W(z_0)f(r) \delta_{tt}.
\end{equation}

Following the above mentioned conditions:
\begin{equation}
    \delta W(z_0) f(r_+,M)+W(z_0) \delta f(r_+,M)=0 .
\end{equation}

The first term is zero due that $f(r_+,M)=0$. On the other hand, since, as it  was above mentioned $W(z) \neq 0$, therefore $W(z_0) \neq 0$. Thus, our problem is reduced to solve $\delta f(r_+,M)=0$:
\begin{equation} \label{espacioDeParametros}
    0 = \frac{\partial f}{\partial r_+} dr_+ + \frac{\partial f}{\partial M} dM .
\end{equation}

Following equation \eqref{espacioDeParametros}, for our four dimensional solution \eqref{funcionf}, where the mass function is given by the equation \eqref{funciondeMasa1}, the first law takes the form:
\begin{equation} \label{duTdS}
    \frac{\partial m}{\partial M} dM = \left ( \frac{1}{4\pi} f'\big |_{r=r_+}  \right ) d \left (\frac{4 \pi r_+^2}{4} \right).
\end{equation}

The above equation can be rewritten as:
\begin{equation} \label{duTdS1}
du=Td \left ( \frac{A}{4} \right ) ,
\end{equation} 
where $A$ corresponds to the area of a three dimensional sphere and  the temperature and entropy are easily computed as: 
\begin{align}
    T&= \frac{1}{4\pi} f'\big |_{r=r_+} \\
    S&=\frac{A}{4} = \pi r_+^2.
\end{align}

 Thus, our computed definition of entropy follow the area's law and so the existence of extra dimensions does not modify the correct values of temperature and entropy at the induced four dimensional geometry. The term $du$ corresponds to a local definition of the variation of the energy defined in \cite{Estrada:2020tbz,Aros:2019auf}. In this definition the factor $dm/dM$ in equation \eqref{duTdS} is always positive, and thus, the sign of the variation of $du$ always coincides with the sign of the variation of the total energy $dM$. Furthermore at infinity it is fulfilled that $\displaystyle \lim_{r \to \infty} dm/dM=1$, therefore the variation of the local and total energy are similar at the asymptotically region, and thus, the first law is reduced to the usual form $dM=TdS$ \cite{Estrada:2020tbz}.

\section{Conclusion and summarize} 

We have provided a new five dimensional black string--like solution. This is based on the embedding of a four dimensional regular black hole solution into a compact extra dimension.

The warp factor depends only on the extra coordinate $W(z)$ and modifies the  four dimensional geometry $d\bar{s}_{4D}^2$, see equation \eqref{elementodelinea}. On the other hand, we have imposed the form of the energy momentum tensor \eqref{EnergiaMomentum}, such that the warp factor also modifies the information of the four dimensional energy momentum tensor \eqref{EnergiaMomentum4D} through the function \eqref{EcuacionW}.

In subsection \eqref{restricciones} we have found relations between the four and five dimensional invariants of curvature. Using this latter, we have enunciated a list of constrains in order to that the complete five dimensional solution to be regular. Following these constraints we have computed the complete five dimensional solution. As example, we have used the Hayward metric as four dimensional toy model. However, following our list, could be constructed another regular black strings-like using other models of four dimensional regular black holes with a de--Sitter core, see for example \cite{Dymnikova:1992ux} . This could be analyzed in elsewhere. 

Due that the four dimensional geometry is regular, instead the formation of a black string where the radial singularity is represented by the product between the Schwarzschild singularity and $S^1$, it is formed a black string-like where the core is represented by the product between the de--Sitter core of the four dimensional Hayward solution and $S^1$. On the other hand, due that the extra coordinate is compact the horizon has topology $S^2 \times S^1$.

Due that the four dimensional solution is asymptotically flat, our complete solution behaves at the infinity of the radial coordinate as an Anti de--Sitter five dimensional space time. On the other hand, also at this asymptote the topology corresponds to the product between Minkowski and $S^1$. Furthermore at this place for zero cosmological constant, $l \to \infty$, our solution behaves as the Kaluza-Klein black string \cite{Gregory:2008rf}.

The induced geometry is characterized by the induced metric at an arbitrary location $z=z_0$. Following local relations, based on the evolution along the space of parameters \cite{Aros:2019auf,Zeng:2019huf}, we have computed the first law of thermodynamics on the induced geometry. So the existence of extra dimensions does not modify the correct values of temperature and entropy at the induced four dimensional geometry.

It is wort to mention that small perturbations around black strings show that these latter are unstable \cite{Gregory:1993vy} and this instability could be stabilized if the extra dimension is compactified to a scale smaller than a minimum value. This issue is outside of the scope of our work, however should be tested in a future work.

\bibliography{mybib}

\end{document}